


\def\brk{\hfill\break}
\def\pps{\vec p_{\perp}^2}
\def\ppsi{\vec p_{i,\perp}^2}
\def\ppi{\vec p_{i,\perp}}
\def\ppj{\vec p_{j,\perp}}

\overfullrule=0pt

 \pubnum{6126}
 \date{March 1993}
 \pubtype{T/E}
 \titlepage
 \title{%
 Parke-Taylor amplitudes in the multi-Regge kinematics
 \doeack}
\author{Vittorio Del Duca}
\SLAC
\abstract{The Parke-Taylor multigluon amplitudes are examined in the
multi-Regge kinematics, which assumes strong rapidity ordering of the produced
gluons, and are used to compute the $n$-gluon production cross section and the
gluon-gluon total cross section.}
\submit{Physics Letters B}
\endpage
\vfill

In hadron scattering processes with large momentum transfer $Q^2$, of
the order of the center of mass energy $s$, we evaluate the physical
quantities of interest by performing perturbative QCD calculations
of parton cross sections as series expansions in the strong coupling
constant $\alpha_s$. Since the calculation of the coefficients in the series
becomes quickly complicated as one goes to higher orders in the
expansion, only very few coefficients are usually computed, that is
calculations are performed at a fixed (and small) order in $\alpha_s$.

At the Tevatron, LHC and SSC hadron colliders a new
kinematical region, the semihard region, characterized by scattering
processes with $s >> Q^2  >> \Lambda^2_{QCD}$, becomes important.
In this region the momentum transfer is large enough to allow
perturbative QCD calculations, but so much smaller than the
center of mass energy that processes with production of a large number
of partons become relevant. In the series expansion of the parton cross
section each coefficient contains the logarithm of a large ratio of
kinematical invariants, of the order of $\ln{(s/Q^2)}$,
and the effective expansion parameter becomes
the product $\alpha_s \ln{(s/Q^2)}$,
which may be $O(1)$. Thus in the series expansion we have to retain
many higher orders, which open up several real channels and are
the cause of the abundant
production of partons. To keep all this into account properly it is
useful to have techniques that resum all the orders in the effective
expansion parameter. To this end it is necessary
to have analytical, albeit approximate, expressions for multiparton amplitudes.

In the semihard region, the leading contribution to scattering
processes always comes from the exchange of a particle of highest
spin, in our case a gluon, in the crossed channel \REF\Azimov{Ya.I.
Azimov, \sl Phys. Lett. \bf 3 \rm (1963) 195.}\refend. In the case of
multiple gluon
emission the rapidity interval between the scattered partons is
filled with gluons. In the multi-Regge kinematics, which yields
the leading logarithmic contribution to the cross section, the
rapidities of the emitted gluons are strongly ordered.
Fadin, Kuraev and Lipatov \REF\FKL{E.A. Kuraev, L.N. Lipatov and
V.S. Fadin, \sl Sov. Phys. JETP \bf 44 \rm (1976) 443.}\refend
computed long ago the multigluon amplitude
in the multi-Regge kinematics. This
amplitude contains all the virtual radiative corrections, whose effect is
to reggeize the gluons exchanged in the crossed channel \refmark\FKL
\REF\Lipatov{L.N. Lipatov, \sl Sov. J. Nucl. Phys. \bf 23 \rm (1976) 338.}
\refend.
Balitsky, Fadin, Kuraev and Lipatov (BFKL) \REF\BFKL{E.A. Kuraev,
L.N. Lipatov and V.S. Fadin, \sl Sov. Phys. JETP \bf 45 \rm (1977) 199; \brk
Ya. Ya. Balitskii and L.N. Lipatov, \sl Sov. J. Nucl. Phys. \bf 28 \rm (1978)
822.}\refend\ computed then the total parton-parton cross section,
by putting the multi-gluon amplitude in the multi-Regge phase space,
integrating out the rapidities of the produced gluons, and reducing the
dependence of the cross section on the gluon transverse momenta to the
resolution of an integral equation. In this equation the infrared real
and virtual divergences exactly cancel and thus the eigenvalues do
not depend on the infrared cutoff. The total parton-parton cross
section is found then to have a power growth with $s$, with the power
depending on the eigenvalue of the integral equation. Via the optical
theorem the total parton-parton cross section is related to the forward
elastic parton-parton scattering with color-singlet exchange in the
crossed channel, the perturbative QCD pomeron.

Another multigluon amplitude, the Parke-Taylor amplitude
\REF\PT{S.J. Parke and T. Taylor, \sl Phys. Rev. Lett. \bf 56 \rm (1986) 2459.}
\refend, which is a tree-level multigluon amplitude in a helicity basis, with
a particular
choice for the gluon helicities, is available in the literature. It is
not specific to a particular kinematical region.
It has been used to make approximate calculations of the four- and
five-jet production rates \REF\KS{Z. Kunst and W.J. Stirling, \sl Phys.
Rev. \bf D37 \rm (1988) 2439.}\refend, which have been found to be in good
agreement with the data \REF\UA{UA2 Collaboration, \sl Phys. Lett. \bf
B268 \rm (1991) 145.}\refend.

In this paper we want to consider the Parke-Taylor multigluon amplitude, for
the
production of an arbitrary number of gluons, in the
multi-Regge kinematics, study the color flows of the produced gluons
on the Lego plot in azimuthal angle and rapidity, and compute the total
gluon-gluon cross section. Since the virtual radiative corrections
are missing in the Parke-Taylor amplitudes, we will have to cut off the
infrared
real divergences and we expect the slope of the pomeron
trajectory to depend on the infrared cutoff.

We will find that introducing an infrared cutoff is not enough, and we
also have to regulate the behavior of the amplitudes in the
ultraviolet, to avoid the rise of an unphysical singularity in the
total cross section. The same happens also in the BFKL multigluon amplitude
if we neglect the contribution of the virtual radiative corrections.
Thus a multigluon amplitude without virtual radiative corrections seems
inherently ill-suited for the calculation of a fully inclusive
quantity, like the total cross section.

\noindent{\bf Parke-Taylor amplitudes in the multi-Regge kinematics}

A tree-level multigluon amplitude can be written in an SU($N_c$)
Yang-Mills theory as

$$ {\cal M}_n = \sum_{[1,2,...,n]'} tr(\lambda^{a_1} \lambda^{a_2}
   \cdots \lambda^{a_n}) \, m(p_1,\epsilon_1; p_2,\epsilon_2;...;
   p_n,\epsilon_n), \eqn\one$$
where $a_1, a_2,..., a_n$, $p_1, p_2,..., p_n$, and $\epsilon_1,
\epsilon_2,..., \epsilon_n$ are respectively the colors, momenta and
helicities of the gluons, $\lambda$'s are the color matrices in the
fundamental representation of SU($N_c$) and the sum is over the noncyclic
permutations of the set $[1,2,...,n]$. The gauge-invariant subamplitudes
$m(p_1,\epsilon_1; p_2,\epsilon_2;...; p_n,\epsilon_n)$ enjoy several
properties\REF\MP{F.A. Berends and W.T. Giele, \sl Nucl. Phys. \bf B294
\rm (1987) 700; \brk M.L. Mangano, S.J. Parke and Z. Xu, \sl Nucl.
Phys. \bf B298 \rm (1988) 653; \brk F.A. Berends and W.T. Giele, \sl
Nucl. Phys. \bf B306 \rm (1988) 759.}\refend,
like incoherence to leading order in $1/N_c$
$$ \sum_{colors} |{\cal M}_n|^2 = N_c^{n-2} \, (N_c^2-1) \sum_{[1,2,...,n]'}
   \bigl[|m(p_1,\epsilon_1; p_2,\epsilon_2;...; p_n,\epsilon_n)|^2 +
   O(N_c^{-2})\bigr], \eqn\two$$
and cyclical and reversal symmetry
$$ \eqalign{m(p_1,\epsilon_1; p_2,\epsilon_2;...; p_n,\epsilon_n) & =
   m(p_2,\epsilon_2;...; p_n,\epsilon_n; p_1,\epsilon_1) \cr
   m(p_n,\epsilon_n; p_{n-1},\epsilon_{n-1};...; p_2,\epsilon_2;
p_1,\epsilon_1)
& =
   (-1)^n m(p_1,\epsilon_1; p_2,\epsilon_2;...; p_n,\epsilon_n).}
   \eqn\rev$$

If we assume that all the gluons are outgoing, the subamplitude
for the maximally helicity violating configuration
$(-,-,+,\cdots,+)$ is given by \refmark\PT

$$ m(p_1^-,p_2^-,p_3^+,...,p_n^+) = i g_s^{n-2} \, {<12>^4 \over
<12><23>\cdots <n1>}\, . \eqn\three$$
where the spinor product is defined as $<pq> = \overline{\psi_-}(p) \psi_+(q)$,
with $\brk \psi_{\pm}(p) = {1 \over 2} (1\pm \gamma_5) \psi(p)$. $\psi(p)$
is a massless Dirac spinor, normalized in such a way that
$|<pq>|^2 = 2(p\cdot q)$.

\FIG\figa{The n-gluon production amplitude.}

By replacing the Parke-Taylor subamplitudes \three\ into
\one\ and using the incoherence to leading order in the number
of colors \two, we obtain the square of the multigluon Parke-Taylor
amplitudes, summed over colors and the maximally helicity violating
configurations, to
leading order in $1/N_c$. From this we straightforwardly derive
the $n$-gluon production squared Parke-Taylor amplitude (fig. \figa),
averaged over colors and helicities of the incoming gluons

$$ \eqalign {& |{\cal M}(p_A,p_0,p_1,...,p_{n+1},p_B)|^2 =
   2 \, {1 \over 4(N_c^2-1)} \, (g_s^2)^{n+2} \, N_c^{n+2} \cr
   & \sum_{i>j} s_{ij}^4 \sum_{[A,0,1,...,n+1,B]'}{1 \over s_{A0} s_{01}
   \cdots s_{n,n+1} s_{n+1,B} s_{AB}}, \cr} \eqn\four $$
where we label the incoming gluon momenta as $p_A$ and $p_B$,
and $\brk i,j = A,0,...,n+1,B$. The overall factor 2 at the beginning of
the right hand side of \four\ is present only in the inelastic
case $n\not=0$, and keeps into account the different
maximally helicity violating configurations (-,-,+,$\cdots$,+) and
(+,+,-,$\cdots$,-).

We parametrize the
momenta of the produced gluons in terms of the rapidity $\eta$, and the
momentum $p_{\perp}$  and the azimuthal angle $\phi$ in the plane
transverse to the beam axis. Then the kinematical invariants are given by

$$ \eqalign{& s_{Ai} = \sqrt{s} \, |\ppi| \, e^{-\eta_i} \cr
   & s_{Bi} = \sqrt{s} \, |\ppi| \, e^{\eta_i} \cr
   & s_{ij} = 2 \, |\ppi| \, |\ppj| \, [cosh(\eta_i - \eta_j) -
   cos(\phi_i - \phi_j)],\cr} \eqn\five $$
where $s = s_{AB}$ is the center of mass energy of the scattering
process and $\brk i,j=0,...,n+1$. Now we want to specify the $n$-gluon
production squared Parke-Taylor amplitude to the semihard regime, where
$s >> m^2  >> \Lambda^2_{QCD}$, and $\pps \simeq m^2$ is the
characteristic value of the transverse momentum of the
produced gluons. To pick up the leading contribution in $ln(s/m^2)$ we
consider the multi-Regge kinematics, where the gluon rapidities are
strongly ordered
$$\eta_A \simeq \eta_0 >> \eta_1 >> \cdots >> \eta_{n+1} \simeq \eta_B.
  \eqn\order $$

In this kinematics the sum over helicities becomes
$$ \sum_{i>j} s_{ij}^4 = 4 \, s^4 \, \bigl(1 + O(s^{-1})\bigr), \eqn\six $$
with $i,j = A,0,...,n+1,B$. If we assume that $s$-channel helicity
conservation for the incoming gluons holds, which is the case in
multi-Regge kinematics\refmark{\FKL,\Lipatov}, then we have $2^{n+2}$
possible helicity configurations in the $n$-gluon production amplitude.
Indeed, fixed the helicity configuration for the incoming gluons,
$s$-channel helicity conservation allows for $2^n$ different helicity
configurations for the $n+2$ outgoing gluons. Then the sum over the
four helicity configurations for the incoming gluons gives the figure
quoted above. Thus from \six\ together with the extra factor 2 in
\four, for $n\not=0$,
due to the different maximally helicity violating configurations \brk
(-,-,+,$\cdots$,+) and
(+,+,-,$\cdots$,-), we see that the Parke-Taylor amplitudes count correctly the
number of helicity configurations for the elastic case $n=0$ and
for the 1-gluon production case.

To study the color ordering of the gluons we introduce the reduced
squared amplitude
$$ |M(p_A,p_0,p_1,...,p_{n+1},p_B)|^2 = s^4 \sum_{[A,0,1,...,n+1,B]'}
{1 \over s_{A0} \, s_{01} \cdots s_{n,n+1} \, s_{n+1,B} \, s_{AB}},
\eqn\nine $$
and the function
$$ F_{ij} = {e^{\eta_i - \eta_j} \over 2[cosh(\eta_i - \eta_j)
   - cos(\phi_i - \phi_j)]}. \eqn\seven $$
In the multi-Regge kinematics $F_{ij}$ becomes
$$ F_{ij} = \cases{1 + O\bigl(e^{-(\eta_i - \eta_j)}\bigr),
   &if \eta_i > \eta_j;\cr
   e^{2(\eta_i - \eta_j)}, &if \eta_i < \eta_j. \cr} \eqn\eight $$

where we assumed that the rapidity interval between any two gluons is
large enough that we can neglect the azimuthal-correlation term in
\seven. Thus in the Parke-Taylor amplitudes in multi-Regge kinematics \order\
the
azimuthal correlation between the produced gluons is not a leading
order effect. We notice, though, that in the BFKL multigluon amplitudes
\refmark\FKL\ there is azimuthal correlation between the produced
gluons, due to the propagators of the gluons exchanged in the $t$
channel. Only at the end of the day, in the solution of the BFKL
integral equation, we do realize that the azimuthal correlation is a
subleading effect.

In considering the sum over colors, we have to sum over all the
non-cyclic permutations of the set [A, 0, 1,..., n+1, B] in \nine.
To do so, let
us fix the position of gluon $A$ in the set and move gluon $B$ one step
at a time to the left, and for each position of gluon $B$ we consider
all the permutations of the $n+2$ outgoing gluons. This will exhaust
all the non-cyclic permutations of the set above. They are not all
different, though, since for
each color ordering in \nine\ there is the reversed ordering
which, because of \rev, yields the same contribution
$$ \eqalign{[A,0,...,m,B,m+1,...,n+1] &= [n+1,...,m+1,B,m,...,0,A]
   \cr &= [A,n+1,...,m+1,B,m,...,0].} \eqn\symm$$

To begin with, let us
consider the color ordering [A, 0, 1,..., n+1, B], plus all the
permutations of the outgoing gluons.
By using the kinematical invariants of \five\ and the function
$F_{ij}$, the reduced squared amplitude \nine\ becomes

$$ |M(p_A,p_0,p_1,...,p_{n+1},p_B)|^2 =  {s^2 \over \prod_{i=0}^{n+1} \ppsi}
   \sum_{\sigma} \prod_{i=0}^n F_{i,i+1}, \eqn\ten $$
   \FIG\one{Multigluon amplitude in the color configuration
   [A, 0, 1,..., n+1, B].}
where $\sum_{\sigma}$ represents the permutations of the $n+2$ outgoing
gluons in the color configuration $[A, 0, 1,..., n+1, B]$, while keeping fixed
the incoming gluons
$A$ and $B$.
In fig.\one\ we represent the color configuration [A, 0, 1,..., n+1, B]
in terms of color lines in the fundamental representation of SU($N_c$).
Permuting the
outgoing gluons in the color ordering [A, 0, 1,..., n+1, B], we see
that the only permutation which respects the strong rapidity ordering
\order\ is the identity $\sigma(i) = i$, all the others giving a
contribution that, because of \eight, is
$O(e^{-|\Delta\eta_{ij}|})$. Thus, to leading order in rapidity,
the reduced squared amplitude \ten\ becomes

$$ |M(p_A,p_0,p_1,...,p_{n+1},p_B)|^2 = {s^2 \over \prod_{i=0}^{n+1}
   \ppsi} \bigl[1 + O(e^{-|\Delta\eta_{ij}|})\bigr]. \eqn\eleven $$

   \FIG\two{(a) Multigluon amplitude in the color configuration \brk
   [A, 0,..., j-1, j+1,..., n+1, B,j], and (b) its untwisted version.}

Now let us take the color ordering [A, 0,..., j-1, j+1,..., n+1, B,
j], where $j=0,...,n+1$ and we consider all the permutations of the
$n+1$ gluons between gluons $A$ and $B$. The squared amplitude is

$$ \eqalign{& |M(p_A,p_0,...,p_{j-1},p_{j+1},...,p_{n+1},
   p_B,p_j)|^2 = \cr & {s^2 \over \prod_{i=0}^{n+1}\ppsi} \sum_{\sigma}
   \sum_{j=0}^{n+1} F_{0,1} \cdots
   F_{j-2,j-1} \, F_{j-1,j+1} \, F_{j+1,j+2} \cdots F_{n,n+1}}. \eqn\twelve $$

This corresponds to the configuration of fig.\two(a), which we untwist
in fig.\two(b). The untwisted diagram can be conventionally thought of
as a double-sided Lego plot in rapidity and azimuthal
angle \REF\bj{G. Veneziano, \sl Phys. Lett. \bf 43B \rm (1973) 413; \brk
J.D. Bjorken, \sl Phys. Rev. \bf D45 \rm (1992) 4077.}\refend.
In this picture $\sum_{\sigma}$ in \twelve\ represents the
permutations of the $n+1$ gluons on the ``front" of the Lego plot.
We notice that the parameters of the gluon on the ``back" of the Lego plot
do not appear in \twelve.
For each gluon that we bring to the back of the Lego plot,
there is one permutation, the identity $\sigma(i) = i$, that gives a
leading contribution to \twelve\ and yields a strong rapidity
ordering of the gluons on the front of the Lego plot. Any other permutation
violates this rapidity ordering and is $O(e^{-|\Delta\eta_{ij}|})$.
Then the leading contribution to \twelve\ is

$$ |M(p_A,p_0,...,p_{j-1},p_{j+1},...,p_{n+1},
   p_B,p_j)|^2 = (n+2) \, {s^2 \over \prod_{i=0}^{n+1} \ppsi} \,
   \bigl[1 + O(e^{-|\Delta\eta_{ij}|})\bigr]. \eqn\thirteen $$

   \FIG\three{(a) Multigluon amplitude in the color configuration \brk
   [A, 0,..., j-1, j+1,..., k-1,k+1,..., n+1, B, k, j], and (b) its
   untwisted version.}

If we move gluon $B$ one more place to the left we have the
color ordering $\brk$ [A, 0,..., j-1, j+1,..., k-1, k+1,..., n+1, B, k, j],
where $j,k = 0,...,n+1$, and we consider all the permutations of the
gluons to the left of gluon $B$, independently from the ones to the
right. The corresponding squared amplitude is

$$ \eqalign{& |M(p_A,p_0,...,p_{j-1},p_{j+1},...,p_{k-1},p_{k+1},...,p_{n+1},
   p_B,p_k,p_j)|^2 = \cr & {s^2 \over \prod_{i=0}^{n+1} \ppsi}
    \sum_{\sigma_F \sigma_B}
   \sum_{j<k} F_{0,1} \cdots F_{j-1,j+1} \cdots
   F_{k-1,k+1} \cdots F_{n,n+1} \, F_{j,k}},
   \eqn\thirt $$
with the configuration of fig.\three(a), and its untwisted version
fig.\three(b), with two gluons on the back of the Lego plot.
$\sum_{\sigma_F \sigma_B}$ represents the permutations of the gluons
on the front and on the back of the Lego plot.
In fig.\three(b), for each two gluons that we bring to the back of the Lego
plot
there is
one permutation of the gluons on the front, the identity $\sigma_F(i) =
i$, that gives a leading contribution to the squared amplitude, and
conversely for each set of $n$ gluons on the front there is the
identical permutation of the two gluons on the back $\sigma_B(i) = i$,
that gives a leading
contribution. So, in order to have the leading order in rapidity,
we must take the identical permutation both on the front
and the back of the Lego plot, which yields a strong
rapidity ordering of the gluons on the two sides of the Lego plot.
Since there are $n+2 \choose 2$ such configurations which respect the strong
rapidity ordering of the gluons on the front and the back of the Lego
plot, the leading contribution to \thirt\ is

$$ \eqalign{& |M(p_A,p_0,...,p_{j-1},p_{j+1},...,p_{k-1},p_{k+1},...,p_{n+1},
   p_B,p_k,p_j)|^2 = \cr \noalign{\smallskip} &
   {n+2 \choose 2} {s^2 \over \prod_{i=0}^{n+1} \ppsi}
   \bigl[1 + O(e^{-|\Delta\eta_{ij}|})\bigr].} \eqn\fourt $$

Then in general, given a color configuration to which corresponds an
untwisted diagram that has $m$ gluons on the back of the Lego plot,
there are $n+2 \choose m$ color configurations which respect the strong
rapidity orderings of the gluons on the front and the back of the Lego
plot and give a leading contribution to the squared amplitude.
Then, to leading order in rapidity, the reduced squared amplitude
\nine\ becomes

$$ |M(p_A,p_0,p_1,...,p_{n+1},p_B)|^2 = 2^{n+2} \, {s^2 \over
   \prod_{i=0}^{n+1} \ppsi}
   \bigl[1 + O(e^{-|\Delta\eta_{ij}|})\bigr]. \eqn\fift $$

{}From \four, \six\ and \fift, we can now write the $n$-gluon
production squared Parke-Taylor amplitude, averaged over colors and helicities
of the incoming gluons, to leading order in rapidity in the multi-Regge
kinematics, as

$$ |{\cal M}(p_A,p_0,...,p_{n+1},p_B)|^2 = 2 \, {(2 g_s^2 N_c)^{n+2} \over
   N_c^2-1} \, {s^2 \over \prod_{i=0}^{n+1} \ppsi} \,
   \bigl[1 + O(e^{-|\Delta\eta_{ij}|})\bigr]. \eqn\sixt $$
As mentioned after \four, the overall factor 2 at the
beginning of the right hand side of \sixt\ is missing in the
elastic case $n = 0$.

We finally notice two
features of the multigluon amplitudes which do not depend on the particular
kinematics chosen: $i)$ in the Parke-Taylor amplitudes there is
interaction only between gluons on the same side of the Lego plot;
$ii)$ the two sides of the Lego plot are indistinguishable. Property $i)$
is hinted in \thirt; property $ii)$ holds because of the reversal
symmetry \rev\ and \symm, which is not peculiar
of the multigluon amplitudes (1) at the tree-level\REF\lance{Lance
Dixon, private communication.}\refend\ and implies that for each
color ordering with $n_F$ gluons on the front
and $n_B$ gluons on the back of the Lego plot there is a color ordering
with $n_B$ gluons on the front and $n_F$ gluons on the back which
yields the same contribution to \nine.

It is clear that also the BFKL amplitude must admit a distribution of
the produced gluons on a double-sided Lego plot. We have not been able,
though, to make such an identification.

\noindent{\bf Parke-Taylor gluon-gluon total cross section}

\REF\MN{A.H. Mueller and H. Navelet, \sl Nucl. Phys.
\bf B282 \rm (1987) 727.} \REF\fad{J.D. Bjorken, \sl Int. J.
Mod. Phys. \bf A7 \rm (1992) 4189.}
We are now in the position to compute the $n$-gluon production and the
gluon-gluon total cross
sections from the Parke-Taylor amplitudes in the multi-Regge kinematics. In
order
to dispose of the infrared divergences we will cutoff the gluon
transverse momenta $p_{\perp}$ at a characteristic scale $m$. This
corresponds to the following experimental setting: in hadron-hadron
scattering, we tag two jets at a large rapidity interval on the Lego
plot and count all the accompanying jets produced in
between, and we require that all the jets have transverse momentum larger
than a cutoff $m$\refmark{\MN,\fad}. The phase space for the
production of $n+2$ gluons in multi-Regge kinematics, where
$s >> m^2$ and $\pps \simeq m^2$ and the gluon
rapidities are strongly ordered \order, can be written as

$$ d\Pi_{n+2} = {1 \over 2s} \, \biggl(\prod_{i=1}^n {d\eta_i \over 4\pi}
   \biggr) \biggl(\prod_{i=0}^{n+1} {d^2 p_{i,\perp} \over (2\pi)^2}
   \biggr) \, (2\pi)^2 \, \delta^2 \bigl(\sum_{i=0}^{n+1} p_{i,\perp} \bigr),
   \eqn\sevent $$
where we have used the conservation of energy and longitudinal momentum to
fix the rapidities of the gluons at the extremes of the Lego plot.

The case where two gluons become collinear, and the related function
$F_{ij}$ \seven\ blows up, is not included in the multi-Regge
kinematics. It may appear, though, as a contribution at the boundary of
the integration in the phase space \sevent. To rule this out, we
strictly enforce the strong rapidity ordering \order, i.e. we assume that
any two gluons cannot get closer in rapidity than a fixed cutoff
$\bar\eta$, defined in such a way that \eight\ is valid over the whole
phase space and the azimuthal correlation is negligible everywhere. The
exact value of $\bar\eta$ is actually unimportant, since in the leading
logarithmic approximation the whole rapidity interval $\eta_A - \eta_B$
is defined up to an additive constant. In the following we will assume
that a cutoff $\bar\eta$ has been introduced and we will neglect it. It
is worth recalling though that the BFKL multigluon amplitudes do not
have such a problem at the boundary of the phase space, since there is
an explicit cancellation of the real and virtual infrared contributions
in the BFKL integral equation.

Using \sixt\ and \sevent, we can compute the $n$-gluon production cross
section. For $n = 0$ we obtain the tree-level elastic cross section, i.e.
the Born term for gluon-gluon scattering

$$ \sigma_{elas} = {8 \over N_c^2 -1} \, {\pi \alpha_s^2 N_c^2 \over 2
   m^2}, \eqn\elas $$
in agreement with ref.\MN. As we see from \sevent, in the $n$-gluon
production cross section
$$ \sigma_n = {1 \over 2s} \int d\Pi_{n+2}
   |{\cal M}(p_A,p_0,...,p_{n+1},p_B)|^2, \eqn\sect$$
the integrals over transverse
momentum are linked by the $\delta$-function. We disentangle them
using the integral representation in impact parameter $b$ space of the
$\delta$-function, and obtain

$$ \int \biggl(\prod_{i=0}^{n+1} {d^2 p_{i,\perp} \over (2\pi)^2}
   \biggr) {1 \over \prod_{i=0}^{n+1} \ppsi} \,
   (2\pi)^2 \, \delta^2 \bigl(\sum_{i=0}^{n+1} p_{i,\perp} \bigr) =
    2\pi \int_0^{\infty} db \, b \left[{K_0(bm) \over 2\pi}\right]^{n+2},
   \eqn\transv $$
where we have introduced the modified Bessel function $K_0$, and the
cutoff $m$ to regulate the infrared behavior of the transverse momentum.
Since $K_0(x)$ is exponentially decreasing for $x >> 1$ and increases
only logarithmically for $x << 1$,
$$ K_0(x) \simeq -\left(\gamma + \ln{x \over 2}\right), \eqn\kappa$$
with $\gamma$ the Euler-Mascheroni constant,
the integral over the impact parameter $b$ is well defined.
The transverse-momentum-conserving $\delta$-function
has provided two more powers of the momentum in the denominator of
the left hand side of \transv, and thus has suppressed
the ultraviolet growth of the transverse
momentum, as expected, since the ultraviolet divergences are an artifact
of the loop corrections and do not appear at tree level.

We perform the integrals over the gluon rapidities, bound by the
rapidity interval $\eta_0 - \eta_{n+1} \simeq \eta_A - \eta_B = \ln(s/m^2)$
between the gluons at the extremes of
the Lego plot, using the strong ordering \order.
Then, fixing $x = bm$ and $z = {\alpha_s N_c \over \pi} \ln(s/m^2)$
and using \transv, the $n$-gluon production cross section \sect\ becomes

$$  \sigma_n = {8 \over N_c^2 -1} \, {2 \pi \alpha_s^2 N_c^2 \over m^2}
    \, {z^n \over n!} \, \int_0^{\infty} dx \, x \, K_0^{n+2}(x).
    \eqn\ncross$$
It shows the growth in rapidity characteristic of the
multi-Regge kinematics. Since the main contribution to the integral over
the impact parameter comes from its lower end, we may use the
approximation \kappa\ for $K_0$. Then the integral over the impact
parameter in \ncross\ has a factorial growth which approximately
compensates the factorial in the denominator, due to the integration in
rapidity. Thus the series \ncross\ becomes geometrical.
Using \elas\ and \ncross, we can write the total cross section for
gluon-gluon scattering as
$$ \sigma_{tot} = {8 \over N_c^2 -1} \, {2 \pi \alpha_s^2 N_c^2 \over m^2}
   \left( \int_0^{\infty} dx \, x \, K_0^2(x) \, e^{z K_0(x)} - 1/4 \right).
   \eqn\total$$
The integral shows an exponential growth in a double logarithm, which
for a large enough $z$ leads to a singularity, i.e. the series
constructed from \ncross\ is not integrable, even though the single
terms \ncross\ in the series are. We remark that
the BFKL total cross section\refmark\BFKL\ does not share such a behavior,
since the
virtual radiative corrections precisely cancel the doubly logarithmic growth
of the real ones, and one is left over with an exponential growth
in rapidity.

Performing the integral over the impact parameter in \total, we obtain
$$ \sigma_{tot} = {8 \over N_c^2 -1} \, {\pi \alpha_s^2 N_c^2 \over m^2}
   \, \left[ e^{-\gamma z} \, {\Gamma(1 - z/2)^4 \over
   \Gamma(2 - z) } \, - {1 \over 2} \right]. \eqn\tot$$
where $\Gamma$ is the Euler gamma function.
When $z \ge 2$, the total cross section becomes singular.\footnote*{Conversely,
we may integrate each term in the
series before resumming it, using the approximation \kappa, and we
obtain $\sigma_{tot} \sim (1 - z/2)^{-3}$ for $z \sim 2$, in agreement
with \tot.} The
singularity comes from the lower end in the integral over the impact
parameter, i.e. from the ultraviolet behavior of the transverse
momenta. That is because the reasoning which follows \transv\ applies
only to finite $n$, but not to the infinite resummation \total.

We notice, though, that the transverse momentum of each of the
produced gluons cannot grow beyond a value $p_{max} <
\sqrt{s}$, because of energy conservation. Thus  we regulate the
integral over the impact parameter in \ncross\ requiring that
$b > 1/p_{max}$, and we evaluate it using \kappa\ and the saddle-point
approximation. As long as $i) \,\, n+2 < 2\, \log{p_{max}\over m}$, the
integral is well approximated by
$$ \int_{m/p_{max}} dx \, x \, K_0^{n+2}(x) \sim (n+2)!\, ,
   \eqn\newa$$
and the $n$-gluon cross section $\sigma_n$ has a geometrical growth; \brk
when $ii)\, n+2 > 2\, \log{p_{max}\over m}$, the integral is better
approximated by
$$ \int_{m/p_{max}} dx \, x \, K_0^{n+2}(x) \sim \left( {m\over
   p_{max}\right)^2 \, \log^{n+2}{p_{max} \over m} , \eqn\newb$$
and the $n$-gluon cross section behaves like
$$ \sigma_n \sim {1 \over p_{max}^2} \, {1\over n!} \, z^n
   \log^{n+2}{p_{max} \over m} . \eqn\newc$$
If $p_{max} = O(m)$, then condition $ii)$ is easily fulfilled and the
$n$-gluon cross section shows a $1/m^2$ behavior, typical of very high
energy cross sections, times an exponential growth in rapidity, in
agreement with the BFKL theory; if $p_{max} = O(\sqrt{s})$, then cross
sections with a small number of gluons exhibit a geometrical growth and
cross sections with a large number of gluons exhibit a $1/s$ behavior,
times an exponential growth in a double logarithm. It looks like
the latter cross sections might give a small contribution to the total
cross section, but, as long as $z > 2$, the exponential growth wins
over the $1/s$ behavior. Indeed the contribution of the $n$-gluon cross
sections, with large $n$, to the total cross section $\sigma_{tot}$ is
$$ \sum_n \sigma_n \sim {1\over m^2}
   \log^2{p_{max} \over m} e^{(z-2) \ln{p_{max} \over m}}, \qquad
   \rm{with} \quad n+2 > 2\, \log{p_{max}\over m} \eqn\asymp$$
which dominates over the contribution of $\sigma_n$ with small $n$ to
$\sigma_{tot}$, as long as $z > 2$. In the BFKL theory, the virtual
corrections suppress the exponential growth in the double logarithm by
not allowing the gluon transverse momenta to become much larger than $m$.

\asymp\ shows the doubly logarithmic growth, typical of a
kinematical regime where there is a strong ordering both in rapidity
and transverse momentum. So even if we have explicitly suppressed this
regime, as said in the discussion which follows \sevent\ since it does
not belong to the multi-Regge kinematics, it does reappear in the total
cross section. It would happen the same in the BFKL total cross section if we
neglected the virtual radiative corrections. Indeed if in the
BFKL integral evolution equation we suppressed the term that
describes the reggeization of the gluon exchanged in the $t$ channel,
i.e. we discarded the virtual radiative corrections, there would be no
cancellation
of the divergences in the eigenvalue of the integral
equation, and in the BFKL total cross section we would have to regulate
by hand the ultraviolet growth of the transverse momentum, obtaining the
same behavior as in \asymp.

This shows an inherent difference between the Parke-Taylor and the BFKL
multigluon amplitudes. Because of the cancellation of the virtual and
real infrared contributions in the BFKL integral equation, the BFKL
amplitudes are particularly suited for the calculation of a fully
inclusive quantity, like the total cross section. For such a quantity
the Parke-Taylor amplitudes do not fare well. They
may be better suited for the calculation of
exclusive quantities, like the $n$-gluon production cross section
\ncross, where the excessive growth due to the lack of virtual
radiative corrections may be taken care of by using appropriate
kinematical cuts, as we have seen above.

\noindent{\bf Acknowledgements}

The author wishes to thank bj. Bjorken, Lance Dixon, Lev Lipatov, Al Mueller
and
Michael Peskin for many stimulating discussions and suggestions,
bj. Bjorken for much encouragement in the investigation of this
problem, and Al Mueller for the saddle-point interpretation of the
$n$-gluon cross section.
\endpage
\refout
\endpage
\figout
\end
\bye